\documentstyle{mn}
\newif\ifAMStwofonts

\ifoldfss
  \ifCUPmtlplainloaded \else
    \NewTextAlphabet{textbfit} {cmbxti10} {}
    \NewTextAlphabet{textbfss} {cmssbx10} {}
    \NewMathAlphabet{mathbfit} {cmbxti10} {} 
    \NewMathAlphabet{mathbfss} {cmssbx10} {} 
  \fi
  \ifAMStwofonts
    \ifCUPmtlplainloaded \else
      \NewSymbolFont{upmath} {eurm10}
      \NewSymbolFont{AMSa} {msam10}
      \NewMathSymbol{\upi}     {0}{upmath}{19}
      \NewMathSymbol{\umu}     {0}{upmath}{16}
      \NewMathSymbol{\upartial}{0}{upmath}{40}
      \NewMathSymbol{\leqslant}{3}{AMSa}{36}
      \NewMathSymbol{\geqslant}{3}{AMSa}{3E}

      \let\leq=\leqslant 
       
    \fi
  \fi
\fi 

\ifnfssone
  \newmathalphabet{\mathit}
  \addtoversion{normal}{\mathit}{cmr}{m}{it}
  \addtoversion{bold}{\mathit}{cmr}{bx}{it}
  \newmathalphabet{\mathbfit} 
  \addtoversion{normal}{\mathbfit}{cmr}{bx}{it}
  \addtoversion{bold}{\mathbfit}{cmr}{bx}{it}
  \newmathalphabet{\mathbfss} 
  \addtoversion{normal}{\mathbfss}{cmss}{bx}{n}
  \addtoversion{bold}{\mathbfss}{cmss}{bx}{n}
  \ifAMStwofonts
    \ifCUPmtlplainloaded \else
      \UseAMStwoboldmath
      \makeatletter
      \new@mathgroup\upmath@group
      \define@mathgroup\mv@normal\upmath@group{eur}{m}{n}
      \define@mathgroup\mv@bold\upmath@group{eur}{b}{n}
      \edef\UPM{\hexnumber\upmath@group}
      \new@mathgroup\amsa@group
      \define@mathgroup\mv@normal\amsa@group{msa}{m}{n}
      \define@mathgroup\mv@bold\amsa@group{msa}{m}{n}
      \edef\AMSa{\hexnumber\amsa@group}
      \makeatother
      \mathchardef\upi="0\UPM19
      \mathchardef\umu="0\UPM16
      \mathchardef\upartial="0\UPM40
      \mathchardef\leqslant="3\AMSa36
      \mathchardef\geqslant="3\AMSa3E

      \let\leq=\leqslant 

    \fi
  \fi
\fi 

\ifnfsstwo
  \DeclareMathAlphabet{\mathbfit}{OT1}{cmr}{bx}{it}
  \SetMathAlphabet\mathbfit{bold}{OT1}{cmr}{bx}{it}
  \DeclareMathAlphabet{\mathbfss}{OT1}{cmss}{bx}{n}
  \SetMathAlphabet\mathbfss{bold}{OT1}{cmss}{bx}{n}
  \ifAMStwofonts
    \ifCUPmtlplainloaded \else
      \DeclareSymbolFont{UPM}{U}{eur}{m}{n}
      \SetSymbolFont{UPM}{bold}{U}{eur}{b}{n}
      \DeclareSymbolFont{AMSa}{U}{msa}{m}{n}
      \DeclareMathSymbol{\upi}{0}{UPM}{"19}
      \DeclareMathSymbol{\umu}{0}{UPM}{"16}
      \DeclareMathSymbol{\upartial}{0}{UPM}{"40}
      \DeclareMathSymbol{\leqslant}{3}{AMSa}{"36}
      \DeclareMathSymbol{\geqslant}{3}{AMSa}{"3E}

      \let\leq=\leqslant 

    \fi
  \fi
\fi 

\ifCUPmtlplainloaded \else
  \ifAMStwofonts \else 
    \def\upi{\pi}
    \def\umu{\mu}
    \def\upartial{\partial}
  \fi
\fi
\title[Gamma-rays from binary system PSR B1259-63/SS2883]
{Gamma-rays from binary system with energetic pulsar and Be star
with aspherical wind: PSR B1259-63/SS2883} 
\author[A. Sierpowska-Bartosik \& W. Bednarek]
       { A. Sierpowska-Bartosik$^{1,2}$ \& W. Bednarek$^{1}$ \\
       (1) Department of Experimental Physics, University of \L \'od\'z,
       ul. Pomorska 149/153, 90-236 \L \'od\'z, Poland \\
       (2) Institut de Ciencies de l'Espai (IEEC-CSIC)
Campus UAB, Fac. de Ciencies, Torre C5, parell, 2a planta
08193 Barcelona, Spain}
\date{Accepted 
      Received ;
      in original form }
\pubyear{}
\begin{document}
\maketitle

\begin{abstract}
At least one massive binary system containing an energetic pulsar, PSR B1259-63/SS2883, has been recently detected in the TeV $\gamma$-rays by the HESS telescopes. These $\gamma$-rays are likely produced by particles accelerated in the vicinity of the pulsar and/or at the pulsar wind shock, in comptonization of soft radiation from the massive star. However, the process of $\gamma$-ray production in such systems can be quite complicated due to the anisotropy of the radiation field, complex structure of the pulsar wind  termination shock and possible absorption of produced $\gamma$-rays which might initiate leptonic cascades. In this paper we consider in detail all these effects. We calculate the  $\gamma$-ray light curves and spectra for different geometries of the binary system PSR B1259-63/SS2883 and compare them with the TeV $\gamma$-ray observations. We conclude that the leptonic IC model, which takes into account the complex structure of the 
pulsar wind shock due to the aspherical wind of the massive star, can explain the details of the observed $\gamma$-ray light curve. 
\end{abstract}


\section{Introduction}

PSR B1259-63/SS 2883 is the best known massive binary system containing a classical radio pulsar in a highly eccentric orbit around the Be type star. The pulsar creates strong relativistic wind which prevents accretion from the wind of the massive star. So then, it is expected that high energy processes, which are characteristic for the isolated radio ($\gamma$-ray) pulsars, can also occur in the inner pulsar magnetosphere of PSR B1259-63. Due to the short rotational period, equal to $47$ ms, and strong surface magnetic field, the pulsar produces a strong relativistic wind which, in contrary to isolated pulsars, interacts with the wind from the massive star. This interaction process is especially important at the periastron passage when the pulsar is immersed in the inhomogeneous massive star wind. As a result of this interaction process, a relativistic shock in the pulsar wind appears. Particles can be accelerated on the shock to $\sim$ TeV energies. Therefore, it has been suspected, that the binary system PSR 1259-63/SS 2883 might produce observable high energy $\gamma$-ray radiation. This expectation has been confirmed during the last periastron passage of the pulsar in 2004, when the HESS detectors discovered the TeV $\gamma$-ray emission from this binary system (Aharonian et al.~2005). 

The production of X-ray and $\gamma$-ray  radiation in PSR B1259-63 system in the model of interaction of the pulsar and stellar winds was considered for the first time  by Grove et al.~(1997), and Tavani et al.~(1997). In those works $e^\pm$ pairs, escaping from the pulsar magnetosphere, are accelerated additionally at the pulsar wind termination shock and interact with the magnetic field and thermal radiation of the massive star. Moreover, the conditions for propagation and cooling of $e^+e^-$ pairs have been considered as a function of the orbital phase. The produced X-ray light curves were in good agreement with the observed X-ray flux variability. The photon spectra were fitted under the assumption that the massive star equatorial wind is inclined to the orbital plain with the inclination angle $i > 25^o$. According to these models, the Lorentz factors of $e^\pm$ pairs in the pulsar wind should be of the order of $\gamma_e \sim 10^6$, while the wind magnetization parameter, which determines the magnetic field strength at the shock, should be $\sigma \sim 10^{-1} - 10^{-2}$  (Tavani et al.~1996). 

A similar model for this binary system has been also considered by Kirk, Ball \& Skjaeraasen~(1999). These authors argued for the first time that also the TeV $\gamma$-rays should be produced in PSR 1259-63/SS 2883 binary system in the IC scattering of soft radiation from the massive star.  Models with domination of the adiabatic and radiative energy losses were considered. However, they predicted the maximum of the TeV emission around the periastron passage which seems to be inconsistent with the observations. However, very recently Khangulyan et al.~(2007) argued that under special conditions such a general leptonic model can explain the TeV $\gamma$-ray light curve observed from this system. Also the hadronic model for production of high energy $\gamma$-rays has been considered by Kawachi et al.~(2004) who postulate acceleration of hadrons at the termination shock in the pulsar wind. In this model the hadrons can pass to the dense equatorial wind of the massive star and produce pions decaying in $\gamma$-rays. The $\gamma$-ray light curve expected in this model shows two maxima corresponding to the situation when the pulsar crosses of the equatorial wind of the massive star and the local minimum at the periastron passage. Such a $\gamma$-ray light curve describes much better the HESS observations. The hadronic model has been also recently discussed by Neronov \& Chernyakova~(2007).

The above mentioned leptonic models have not taken into account the influence of the effects connected with an aspherical wind of the Be star. Such stars are characterized by dense but slow equatorial winds and low density but fast poloidal winds. The pulsar orbiting around the massive star has to pass trough these different winds. When it is immersed in  the equatorial wind, the shock appears relatively close to the neutron star surface. In contrast, in the polar wind, the shock extends to relatively close to the massive star surface. In this last case, particles accelerated at the shock propagate through a much stronger radiation field and the ICS process should occur more efficiently. Moreover, the $\gamma$-rays produced close to the stellar surface can also be partially absorbed, initiating in this way the IC $e^{\pm}$ pair cascade. The importance of the $\gamma$-ray production in the cascades occurring close to the surface of the massive star has been considered in the past, but only in the case of the isotropic winds (e.g. Sierpowska \& Bednarek 2005a). In this paper, we consider those leptonic processes in a much more complicated geometry, i.e. by applying the aspherical wind properties. We calculate the $\gamma$-ray spectra and light curves expected in such a more complicated and realistic model. The results of our calculations are compared with the recent observations of this binary system by the HESS telescopes in the TeV $\gamma$-ray energies. The preliminary results has been published in Sierpowska \& Bednarek~(2005b) and more details of the calculations are included in the PhD thesis of Sierpowska-Bartosik (Sierpowska-Bartosik~2006).

\section{The binary system PSR B1259-63/SS2883}
\begin{figure}
  \vspace{7.5truecm}
  \includegraphics{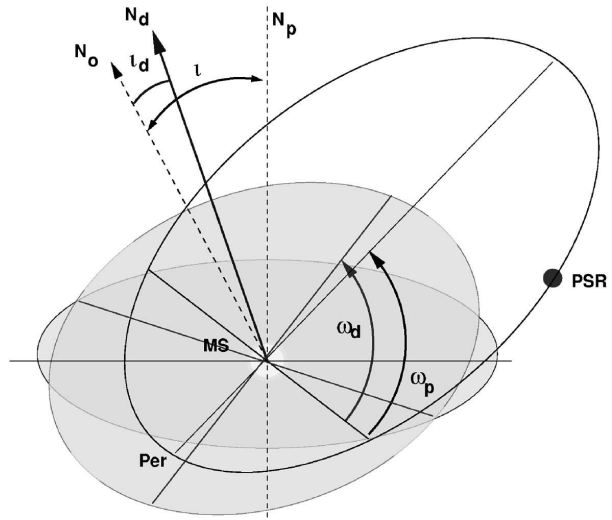}
  \caption{Definition of the basic parameters which describe the complex geometry of the binary system.  $N_{\rm p}$ denotes direction towards the observer, $N_{\rm o}$ is the normal to the orbital plane and $i$ is the inclination angle of the orbital plane. The inclination of the massive star equatorial wind plane to the orbital plane is  denoted  by $i_{\rm d}$, while $N_{\rm d}$ is the normal to the plane of the massive star equatorial wind. The longitude of the periastron passage and the longitude of the plane of the equatorial wind with respect to the periastron passage (defined in the same way) are marked by $\omega_{p}$ and $\omega_{d}$, respectively.} 
\label{fig1}
\end{figure}

\begin{quote}
\begin{table}
  \begin{tabular}{ll}
\hline the orbit & massive star SS2883/Be \\
\hline
$a = 5\rm AU (=177\, R_{Be})$ & $R_{Be}= 6 R_{\odot}$ \\
$\varepsilon = 0.87$ & $M = 10\ M_{\odot}$\\
$i=36^o$ & $\dot{M}\sim2\times10^{-7}\,M_{\odot}\,\rm yr^{-1}$\\
$P_{orb} = 1236.72\,\rm days$ & $V \sim 2\times10^8 \,\rm  cm\,\rm
s^{-1}$\\
$d = 1.5-4.5 \,\rm  kpc$ & $L = 3.3\times 10^{37}\,\rm  erg\,\rm
s^{-1}$\\
 &   $T_{\star}=2.7\times 10^4\,\rm  K$\\
\hline
pulsar PSR 1259-63 & \\
\hline
$P = 47.7\,\rm  ms$ & \\
$L = 8.3\times 10^{35}\,\rm  erg\,s^{-1}$  & \\
$B=3.3\times10^{11}\,\rm G$ & \\
\hline
\end{tabular}
  \caption{Parameters of the PSR B1259-63/SS2883 binary system.}
  \label{tab1}
\end{table}
\end{quote}

The pulsar PSR B1259-63 was discovered during a radio survey of southern Galaxy plain in 1992 (Johston et al. 1992a). It has a period of $P = 47\rm ms$  and period derivative $\dot{P} = 2.28 \times 10^{-15}$. The surface magnetic field has been estimated to $B_{\rm NS} \approx 3.3 \times 10^{11}\,\rm G$ based on the above values and the rotating magnetic dipole model. The pulsar is in an orbit around a massive star of Be type with eccentricity $e=0.87$ and period 1236.72 days (Johnston et al. 1992b, 1994). The system is viewed by the observer at an inclination angle $\sim 36^o$ (Wex et al. 1998). The separation between the stars changes within the range $(0.65-9.3)\,\rm AU$ ($(0.97-14.0)\times 10^{13}\,\rm cm$). At the periastron the pulsar is at the distance of $\sim$23 R$_\star$, while at the apastron at $\sim$177 R$_\star$ (the radius of the Be star is $R_\star = 4.2\times 10^{11}$ cm). The effective temperature of the massive star (B2e) is $\sim 2.7\times 10^4 \,\rm  K$ (Tavani et al. 1997). The distance to the binary system is not precisely known, as from the photometric method it is about $d\sim$1.5 kpc  (Johnston et al. 1996), while from the radio dispersion measure $\sim$4.5 kpc (Taylor \& Cordes 1993). The basic parameters of this binary system are listed in Table 1. Due to the presence of an aspherical massive star wind, it is expected that the pulsar passes through the equatorial wind at phases around $\sim (\tau-13 \pm 5)$ days before periastron (noted as $\tau$) and after periastron, within $\sim (\tau+17 \pm 5)$ days (Connors 2002). The geometry of the binary system and definitions of the basic orbital parameters are shown in Fig. 1.

\subsection{High energy observations}

The binary system PSR B1259-63 was observed in X-rays by the ROSAT and ASCA experiments. The measured luminosity after the periastron passage was $L_x = (0.8-35) \times 10^{33}\, \rm erg\, s^{-1}$ (assuming d = 2kpc, Cominsky et al. 1994). Detection of X-rays together with the radio emission implied that the accretion of matter on the compact object does not occur. Observations at the same periastron passage have been also performed by the Compton Gamma Ray Observatory (Grove et al. 1995, Tavani et al. 1996). The OSSE detector discovered hard X-ray emission with a flat spectrum $dN_{\gamma}/dE = (2.8 \pm 0.7) \times 10^{-3} (E/100 {\rm keV})^{-\alpha} \rm cm^{-2} s^{-1} MeV^{-1}$ and spectral index $\alpha = 1.8 \pm 0.6$ (Grove et al. 1995). At higher energies, only the upper limits could be set (Tavani 1996). During the last periastron passage the binary system has been also observed by the INTEGRAL satellite in energy range $~(3 \rm keV - 10 \rm MeV)$ (Shaw et al. 2004). \\

The TeV observations by the CANGAROO telescopes resulted only with upper limits at the level of $~13\%$ of Crab flux at $\sim$47 and $\sim$157 days after periastron (Kawachi et al. 2004). Close to the periastron, the TeV $\gamma$-ray emission was detected by the HESS telescopes (Aharonian et al. 2004). The differential energy spectrum has been described by a simple power law, $dN/dE = F_0 (E/1\rm TeV)^{-\alpha}$, with the spectral index $\alpha = 2.7 \pm 0.2_{stat}$ and $F_0 = (1.3 \pm 0.1_{stat}) \times 10^{-12}\rm cm^{-2} s^{-1}$.  The spectral index was stable with the phase of the binary system but the level of emission was clearly variable in the range $F_0 = (1-2) \times 10^{-12}\rm cm^{-2} s^{-1}$. The integral average photon flux is $F(>380\rm GeV) = (4.0\pm0.4) \times 10^{-12}\rm cm^{-2} s^{-1}$ corresponding to $~4.9 \%$ of the Crab units. The maximum TeV emission was detected after the periastron at the phase $\sim \tau+15$ days. After that phase, the flux decreases up to the phase $\tau+75$ days. HESS observations did not cover phases just around periastron passage because of bad atmospheric conditions. The average photon flux above $380$ GeV corresponds to a source luminosity $L\approx 8 \times 10^{32} \rm erg s^{-1}$ (at the distance $d=1.5$ kpc) which is only  $\sim$0.1$\%$ of the rotational energy loss rate of the pulsar PSR B1259-63.

\subsection{The structure of the massive star wind}

The observations of the massive star SS2883 (IR and UV energy range) indicate the complex nature of the stellar wind as expected in the case of the Be type star winds (Waters, 1988). They are composed with an equatorial and polar winds characterized by different density and wind velocity. The interpretation of the presence of the equatorial wind in this binary system is also supported by the radio observations. Pulsed radio emission is eclipsed for about $~5$ weeks around periastron ($\sim \tau-16, \tau+15$ days) and it is depolarized for about $~200$ days after the radio eclipse period. These observations are interpreted in terms of the model in which the plane of the dense equatorial wind is inclined to the orbital plane of the system (Melatos 1995, Wex 1998). Based on such a model the pulsar passes twice through the equatorial wind, before the periastron at the phases $(\tau-13 \pm 5)$ days and after it at the phases $(\tau+7 \pm 5)$ days (Connors 2002).

The equatorial stellar wind is dense and has relatively low velocity. On the other hand, the polar wind has much higher velocity but its density is much lower is respect to the equatorial region. The polar wind velocity can be approximated by the profile (Casinelli 1979),
\begin{equation}
V_p(r) = V_0 + (V_{\infty} - V_0)(1-R_\star/r)^{\beta},
\label{eq1}
\end{equation}
\noindent
where $V_{\infty}$ is the terminal velocity of the wind at infinity, $V_{\rm 0} = 0.01 \, V_{\infty}$ is the initial velocity, and the index is usually taken equal to $\beta = 1.5$. For the Be type star one may assume that $V_{\infty} \approx (1-2)\times \, 10^3 \,\rm km\,\rm s^{-1}$. The equatorial wind  velocity is described by, 
\begin{equation}
V_d(r) = V_0 (r/R_\star)^{1.25}, 
\label{eq2}
\end{equation}
\noindent
where the initial velocity is of the order of $V_{\rm 0} = 10\,\rm km\,\rm s^{-1}$ (see e.g. Lamers 1987). As a result the mass loss rate in the polar region, $\dot{M}_{\rm p}$, is much lower than that one in the equatorial region $\dot{M}_{\rm d}$. For the Be type star, the relation between the mass loss rates of these two regions is $\dot{M}_{\rm p} = (10^{-2} - 10^{-4}) \dot{M}_{\rm d}$. The interaction of the pulsar wind with such a complex stellar wind is additionally complicated by the elongated and inclined orbit of the pulsar.

\subsection{Interaction of the pulsar and stellar winds}

\begin{figure*}
  \vspace{6truecm}
  \includegraphics{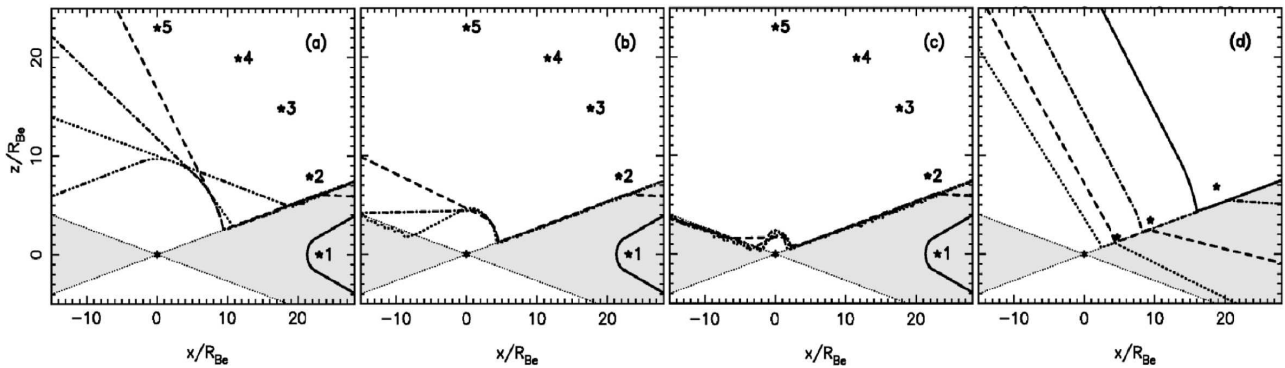}
  \caption{
Formation of the termination shock inside the binary with the separation $r_{\rm p} = 23 R_{\star}$ for different inclinations of the massive star equatorial wind $i_{\rm d}$. The XZ plane is perpendicular to the massive star equatorial wind plane with the star in the center. The opening angle of the equatorial wind is fixed to $\theta_{\rm d} = 30^{\rm o}$ (shaded area). In Fig. (a) the ratio of the mass loss rate of the polar to equatorial wind is $\dot{M}_{\rm p}/\dot{M}_{\rm d} = 10^{-2}$, in (b) $10^{-3}$, and in (c) $10^{-4}$, while the pulsar spin-down luminosity is from Table 1. The termination shock formed for different inclination is plotted with specific lines. The corresponding pulsar position is marked by the number: (1, solid line) i$_{\rm d} = 0^{\rm o}$, (2, dashed line) i$_{\rm d} = 20^{\rm o}$, (3, dot-dashed line) i$_{\rm d} = 40^{\rm o}$, (4, dotted line) i$_{\rm d} = 60^{\rm o}$, and (5, dot-dot-dot-dashed line) i$_{\rm d} = 90^{\rm o}$. (d) The dependence of the shock structure on the binary separation. The position of the  termination shock for the pulsar distance from the massive star r = 5 R$_\star$ is plotted with the dotted line, for r = 10 R$_\star$ with dashed line, for r = 20 R$_\star$ with dot-dashed line and for r = 40 R$_\star$ with solid line. Other wind parameters are: the equatorial wind opening angle $\theta_{\rm d} = 30^o$, its inclination i$_{\rm d} = 20^o$ and the mass loss rate $\dot{M}_{\rm p} = 10^{-3} \dot{M}_{\rm d}$.}
\label{fig2}
\end{figure*}

The observations of PSR B1259-63/SS2883 indicate that the rotational axis of the massive star is inclined to the orbital plane of the binary system. Therefore, the pulsar passes trough wind regions with different properties. Assuming that the pulsar power is constant with the orbital period, it is expected that the termination shock is closer to the massive star surface when the pulsar is immersed inside  the polar wind and it is much closer to the pulsar surface when the pulsar is inside the equatorial wind. The geometry of the termination shock has to be quite complex when the pulsar is close to the transition region between these two types of winds. Moreover, the wind cavity around the pulsar is not axially symmetric  with respect to the axis defined by the stars in contrary to the case spherically symmetric winds (Girard \& Wilson 1987). In the case of the spherical winds, the geometry of the shock in the binary system with a pulsar of power, $L_{\rm PSR}$, is determined by the parameter
\begin{equation}
\eta=L_{\rm PSR}/(c\dot{M}_{\rm Be} V_{\rm Be}), 
\label{eq3}
\end{equation}
\noindent
where $\dot{\rm M}_{\rm Be}$ and  $V_{\rm Be}$ are the massive star wind parameters equal to  $\dot{\rm M}_{\rm p}$ or $\dot{\rm M}_{\rm d}$ and $V_{\rm p}$ and $V_{\rm d}$, depending on the type of the wind (polar or equatorial, respectively).

The case of the wind considered here is much more complicated even if we assume, as the first approximation, that the pulsar wind is spherically symmetric while the massive star wind has the complex structure discussed above. The equatorial wind region can be defined by the opening angle $\theta_d$, its inclination $i_d$ to the line of sight and the longitude prior to the periastron $\omega_d$ (with respect to the orbital plane). We find the localization of the shock front by comparing the wind pressures given by the velocity and the mass loss rates. The geometry and localization of the termination shock depend on the pulsar phase and on the localization of the stellar equatorial wind with respect to the orbital plane. The pulsar phase is denoted as $\varphi$ (starting from the periastron) and is given by the orbital parameters, the longitude of periastron $\omega_p$ and the binary inclination $i$. It is useful to introduce the parameter $\Delta \omega = \omega_d - \omega_p$ (see Fig.~1). If the pulsar passes through the massive star equatorial plane at phases $\varphi = \pm 90^{o}$ this parameter is $\Delta \omega = 0$. 

The specific geometry of the termination shock for different parameters of the stellar wind, but for the same orbital parameters, are shown in Fig. ~\ref{fig2}. Note, that in the region of dense and slow equatorial wind, $\eta$ is lower than unity and the shock front is closer to the pulsar (see e.g. the case marked by '1' in Fig. ~\ref{fig2}(a, b, c). On the other hand, for the polar wind, $\eta$ is larger than unity and the shock front is closer to the massive star surface. For that reason, when the pulsar is immersed inside the poloidal wind, the shock surface can appear quite close to the massive star. Leptons accelerated at these parts of the shock propagate in the strong radiation field of the massive star and they can comptonize stellar photons very efficiently.
\section{Model for high energy radiation}
The complicated shock structures, which can appear in the model taking into account the different proprieties of the poloidal and equatorial stellar winds, provides a good opportunity for efficient production of high energy radiation. However, the presence of the strong (or weak) radiation field can influence the conditions for acceleration of leptons at the relativistic shock due to the radiation energy losses. All these processes, the acceleration, production of photons and their possible absorption, should essentially depend on the location of  the shock, i.e. on the parameters and geometry of the binary system. Below we consider these important processes.

\subsection{Acceleration of leptons} 
Leptons can be already accelerated in the inner pulsar magnetosphere in the electric field 
induced by rotating magnetic dipole. These leptons initiate cascades in the radiation and/or magnetic field as considered in different models (polar gap e.g. Ruderman \& Sutherland~1975, Daugherty \& Harding~1982, outer gap e.g. Cheng, Ho, Ruderman~1986, slot gap, e.g. Arons \& Scharlemann~1979). The charged products of these cascades escape through the light cylinder. They can be farther accelerated in the pulsar wind zone (Kennel \& Coroniti~1984). Here we consider two possibilities for the spectrum of primary leptons in the pulsar wind zone (PWZ) which can represent the limiting cases. In the first model (Model A), we assume that leptons inside the PWZ have a power-law spectrum with  index -1.2 in the energy range $500\, \rm MeV - 500\, \rm GeV$ as they are injected through the light cylinder (see Hibshman \& Arons~2001). In the second model (model B), leptons are monoenergetic with energies $E_e = 10^7\,\rm MeV$ (Kennel \& Coroniti~1984).

The leptons which reach the shock region can be also accelerated in the classical shock acceleration scenario. However, the details of this acceleration process are complicated since, as we showed above, the shock structure can be quite complex. In order to estimate the maximum energies of accelerated leptons, we have to determine the magnetic field strength at different parts of the shock.  Note, that the magnetic field strength at different parts of the shock also change significantly with the orbital phase. The acceleration time of leptons at the specific location at the shock depends on the magnetic field strength from the pulsar side. It can be estimated by,
\begin{eqnarray}
\tau_{acc}\approx (R_{\rm L}/c)(c/V_{\rm s})^2, 
\label{eq4}
\end{eqnarray}
\noindent
where the Larmor radius of leptons is $R_{\rm L} = e m c^2 \gamma/B(R_{\rm s}) \approx 100 \gamma r_{\rm s}/\sigma^{1/2} \rm cm$, $B(R_{\rm s}) =  \sqrt{\sigma} B_0 R^3_{PSR} R^{-2}_{LC} r^{-1}$ (the magnetic field in the termination shock is compressed and finally $B_s = 3\, B_{PSR}(r_s^{PSR})$ (Khangoulian \& Bogovalov, 2003)), $V_{\rm s}$ is the velocity of plasma in the termination shock, and $r_{\rm s}$ is the distance from the pulsar to the termination shock. For ultra-relativistic shock waves, the velocity of plasma in the frame of the shock approaches $V_{\rm s} \sim c/3$, (Achterberg et al. 2001). Then, $\tau_{\rm acc} \approx {9 R_{\rm L}}/{c} \approx 3\times 10^{-8}\gamma r_{\rm s}/{\sigma}^{1/2}~~{\rm s}$. On the other hand, leptons lose energy on radiation processes. The synchrotron cooling timescale is,
\begin{eqnarray}
\tau_{\rm syn} =  (m c^2 \gamma)/(\sigma_{\rm T}cU_{\rm B}\gamma^2),
\label{eq5}
\end{eqnarray}
\noindent
where $\sigma_{\rm T}$ is the Thomson cross section,  $U_{\rm B} = B^2 / (8\pi)\approx 2.5 \times 10^{10} B^2\, \rm eV\, cm^{-3} = 5.8 \times 10^{12} \sigma / r_{\rm s}^2 \, \rm eV\,\rm cm^{-3}$ is the energy density of the magnetic field, where $B$ is in Gauss. For the shock in PSR B1259-63/SS2883, we get approximation  $\tau_{\rm syn} \approx  3.3 \times 10^6 \, {r_{\rm s}^2}/{\sigma \gamma} ~~{\rm s}$. The inverse Compton cooling time scale in the stellar radiation (in the Thomson regime) can be estimated by,
\begin{eqnarray}
\tau_{\rm IC}^{\rm T} =  (m c^2 \gamma)/(\sigma_{\rm T} c U_{\rm rad}\gamma^2),
\label{eq6}
\end{eqnarray}
\noindent
where $U_{\rm rad} = 4 \sigma_{\rm SB}T_{\star}^4 r_{\rm s/Be}^{-2}/c$ is the energy density of the radiation field given by the surface temperature of the massive star $T_{\star}$. For a temperature of the  massive star equal to $T_{\star}=2.7\times10^4\,\rm K$ and at the distance from its surface 
$R_{\rm s}$, $U_{\rm rad} \approx 4.7 \times 10^{-3} T_{\star}^4 r_{\rm s/Be}^{-2} \, \rm eV\,\rm cm^{-3}$, where $r_{\rm s/Be} = R_{\rm s} /R_{\rm Be}$ is the distance from the surface of the massive star to the shock.  In the Klein-Nishina (KN) regime ($\gamma_{\rm T/KN}\gg mc^2/3k_{\rm B} T_{\star}$) the cooling time can be approximated by applying the Lorentz factor of lepton $\gamma_{\rm T/KN}$ at the transition between the T and KN regimes. In the KN regime energy losses on ICS depend only logarithmically on the electron Lorentz factor, $\tau_{\rm IC}^{\rm KN}(\gamma) \approx 10^{-10} \, \gamma^2\, \tau_{\rm IC}^{\rm T}$. For the considered massive star, we obtain
\begin{eqnarray}
\tau_{\rm IC}^{\rm KN}(\gamma) \approx 7.7 \times 10^{-7} \, \gamma r_{\rm s/Be}^2
\rm~~{s}.
\label{eq7}
\end{eqnarray}
The maximum energies of leptons accelerated at the shock are obtained from comparison of the energy gained by them with energy losses. Provided that  the synchrotron losses dominate, the maximum energies of leptons are,
\begin{eqnarray}
\gamma^{\rm max}_{\rm syn}\approx 10^7 r_{\rm s}^{1/2}\sigma^{-1/4}.
\label{eq8}
\end{eqnarray}
\indent
If the acceleration is balanced by IC losses in the Thomson regime, $\gamma\ll  mc^2/ 3k_{\rm B}T_\star$, the maximum Lorentz factors of leptons are 
\begin{eqnarray}
\gamma^{\rm max}_{\rm ICS}\approx 5.1 \times 10^5r_{\rm s/Be}\sigma^{1/4}r_{\rm s}^{-1/2}.
\label{eq9}
\end{eqnarray}
The acceleration process of leptons can be also limited by the time they spend in the acceleration region. This time scale is defined by the advection process of leptons along the termination shock. The advection time can be estimated by dividing the characteristic distance scale of the system by the velocity of the plasma after the shock (equal to $\sim c/3$),
\begin{eqnarray}
\tau_{adv} \approx R_{\rm s}/{V_{\rm s}} \approx 42  \,  r_{\rm s}~~{\rm s}.
\label{eq10}
\end{eqnarray}
\noindent
However, this last process becomes important only for leptons with Lorentz factors $\gamma^{\rm max}_{\rm adv} \approx 1.4\times 10^9  \sqrt{\sigma}$ (from comparison of Eq.~\ref{eq4} and~\ref{eq10}). 

It is usually argued, that, as a result of acceleration at the relativistic shock, leptons obtain a power-law spectrum with spectral index, $\alpha_i$. For strong relativistic shocks, the spectral index is close to, $\alpha_i \sim 2.2 - 2.3$ (see recent models by Achterberg et al.~(2001) or Ostrowski \& Bednarz~(2002)). The normalization factor in the spectrum depends on the acceleration efficiency. In the case of the Crab Nebula it is the order of $\sim 10\%$. 

In the regions of the shock which are relatively close to the surface of the massive star, the radiation field is strong but the magnetic field is relatively weak due to larger distance from the neutron star surface. We estimate the maximum energies of locally accelerated leptons at the shock by taking into account both, their synchrotron and IC energy losses as determined above.

\subsection{Escape of gamma-rays}
\begin{figure}
  \vspace{13truecm}
  \includegraphics{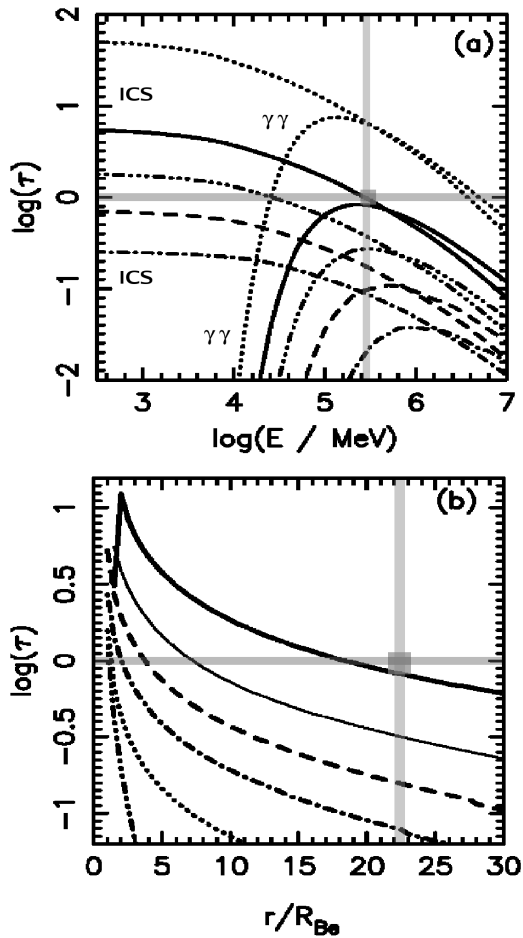}
  \caption{The optical depths, $\tau$, for leptons to Inverse Compton Scattering
(ICS) and for $\gamma$-rays to the $e^\pm$ pair production process in interactions with soft photons from the massive star. The surface temperature of the star SS 2883 is $T_{\star} = 2.7 \times10^4\,\rm K$ and its radius is $R_\star = 4.2\times 10^{11}$ cm. (a) The optical depths as a function of energy of the interacting particle, assuming their rectilinear propagation from the injection place, at $r \approx 23 R_\star$, up to infinity (except for the angle $\theta=180^o$ which is the direction towards the center of the massive star center). Specific curves show the results corresponding to different angles of the injection: $\theta = 60^o$ (dot-dashed line), $\theta = 90^o$ (dashed), $\theta = 120^o$ (dot-dot-dot-dashed), $\theta = 150^o$ (solid), $\theta = 180^o$ (dotted). (b) The optical depths for leptons to ICS process as a function of the distance from the surface of the massive star. The energy of leptons is fixed to $E_e = 4\times10^5\,\rm  MeV$. The angles of injection are the following: $\theta = 0^o$ (dot-dot-dot-dashed  line), $\theta = 30^o$ (dotted), $\theta = 60^o$ (dot-dashed), $\theta = 90^o$ (dashed), $\theta = 120^o$ (thin solid), $\theta = 150^o$ (thick solid). The gray square marked in the both plots shows the value of the optical depth equal to unity for leptons with energy $E_e = 4\times 10^5\,\rm  MeV$, injection angle $\theta \sim 150^o$, and at a distance $r \sim 23 R_\star$.}
\label{fig3} 
\end{figure}
For some phases of the pulsar in its orbit, leptons can be accelerated within a few radii from the surface of the massive star. TeV $\gamma$-rays produced at such distances can be absorbed in the radiation field  of the massive star. As a result, an IC $e^\pm$  pair cascade in the anisotropic radiation of the massive star can develop as considered e.g. by Bednarek (1997) or Sierpowska \& Bednarek (2005a). The conditions for the propagation of leptons and $\gamma$-ray photons in the radiation field of the massive star can be analyzed from the optical depths for leptons on the IC scattering and for $\gamma$-rays on  $e^\pm$ pair production. To simplify the problem we assume that these particles propagate linearly from the place of injection at the distance from the center of the massive star, $x_i$, and at fixed angle, $\theta_{obs}$, measured with respect to the direction defined by the centers of the stars.  The results of these calculations are shown in Fig. ~\ref{fig3}. It is clear, that leptons and $\gamma$-rays (with energies above a few $10^5$ MeV) injected within the distance of the periastron passage of PSR B1259-63 and propagating within $60^{\rm o}$ towards the massive star should interact frequently (Fig. ~\ref{fig3}a shows the results for $r = 23 R_\star$). The optical depths to ICS increase significantly for smaller distances from the massive star (see Fig. ~\ref{fig3}b). Therefore, we conclude that the $\gamma$-ray absorption process has to be also taken into account even in the case of not so compact binary system as PSR B1259-63/SS2883.

\subsection{Gamma-ray production}
\begin{figure}
  \vspace{6.truecm}
  \includegraphics{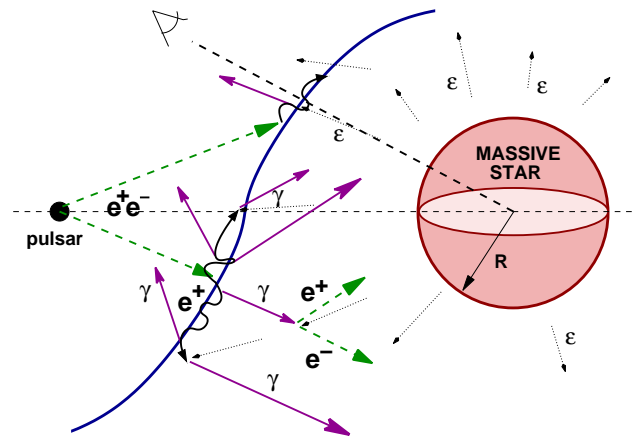}
  \caption{The scenario for the gamma-ray production process inside the binary system with complex structure of the termination shock. $e^+e^-$ pairs escaping from the pulsar light cylinder propagate through the pulsar wind zone. They are additionally accelerated in the pulsar wind termination shock formed by the interaction of the pulsar and stellar winds.  Gamma-rays are produced by leptons due to the Inverse Compton scattering of stellar radiation  from the massive star. Those photons which propagate close to the stellar surface can be  absorbed initiating secondary generation of $e^\pm$ pairs. As a result, IC $e^\pm$ pair cascade can develop provided that a part of the shock structure is relatively close to the surface of the massive star.} 
\label{fig4}
\end{figure}

Injected leptons start to produce $\gamma$-rays (by IC scattering of stellar photons) already inside the pulsar wind zone (PWZ), after that, during their propagation inside the shock region, and finally in the cascade process during propagation of $\gamma$-rays in the  massive star wind region (MSWR). However, the relative contribution of these regions to the $\gamma$-ray spectrum escaping to the observer depends on the distance of the pulsar from the massive star (the phase of the binary system) and on the location of the shock.

Let us consider at first the pulsar wind zone (PWZ). It is assumed that leptons propagate rectilinearly in this region. Produced by them $\gamma$-rays pass through the shock and some of them might be absorbed in the region dominated by the wind of the massive star. In this region, only leptons propagating directly towards the observer can produce observable $\gamma$-rays.
 
Leptons which were able to reach the shock region are farther accelerated in the shock acceleration mechanism. They obtain the power law spectra with maximum energies determined by the escape or radiation processes (see Sect.~3.1). We assume that leptons at the shock are isotropised in the rest frame of the plasma moving with the velocity  $v_{\rm adv} = c/3$ along the shock surface. These leptons produce $\gamma$-rays in the IC scattering of stellar radiation. The details of this process are quite complicated since the radiation field is not isotropic with respect to the location of leptons at the shock. Therefore, the efficiency of $\gamma$-ray production and their spectra depend on their escape direction (measured with respect to the massive star). Part of produced $\gamma$-rays (mainly those moving in the general direction
towards the massive star, i.e. through the massive star wind region) can be absorbed initiating  the IC $e^\pm$ pair cascade. Also the synchrotron energy losses of these secondary (and primary leptons) are taken into account in our calculations. The schematic presentation of processes discussed here are shown in Fig.~\ref{fig4}. As a final result, we obtain the $\gamma$-ray spectra escaping from the binary system at an arbitrary location of the observer and different phases of the compact object. 

\section{$\gamma$-rays from PSR 1259-63/SS 2883}
\begin{figure*}
  \vspace{8.5truecm}
  \includegraphics{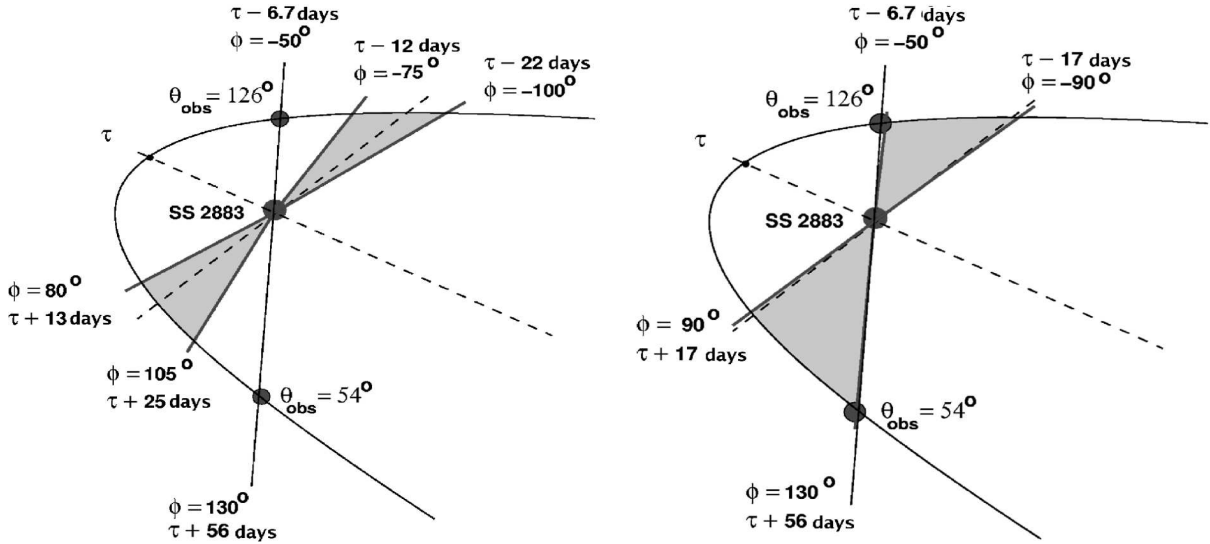}
  \caption{The geometry of the binary system for specific massive star
equatorial wind in model I ($\Delta\omega_d=5^o,\, \theta_d=20^o,\, i_d=40^o$,
left figure) and model II ($\Delta\omega_d=20^o,\, \theta_d=20^o,\, i_d=30^o$, right figure). The shaded areas correspond to the phase ranges when the pulsar crosses the massive star equatorial wind.}
\label{fig5}
\end{figure*}

Low energy observations of the system indicate that the plane of the equatorial wind is inclined with respect to the orbital plane of the pulsar. This angle is estimated in the range $i_d \sim (10^o, 40^o)$. The opening angle of the equatorial wind is  limited to the range $\theta_d \sim (5^o, 40^o)$. 
Because the inclination of the binary system with respect to the observer is $i = 36^o$, the range of observation angles, which determine the IC $e^{\pm}$ pair cascade process, is limited to $54^o \leq \theta_{obs} \leq 126^o$, where $\theta_{obs}$ is the angle between the axis defined by the stars and the direction to the observer. The parameters of the wind of the massive star SS2883 are not well known. Therefore, we have to cover specific range of wind velocities and mass loss rates which are expected to be typical for the Be type stars. 

The periods when the pulsar passes through the equatorial wind are determined from the radio observations (Connors et al. 2002). We analyze different sets of parameters describing the geometry of the relative location of  the equatorial disk with respect to the orbital plane which give similar periods of wind disc crossing by the pulsar. The parameters which have to be fixed for every model,   $\theta_d$, $i_d$, and $\omega_d$, are defined in Fig.~\ref{fig1}. Observations of the binary system PSR B1259-63/SS288 suggest that the plane of the massive star equatorial wind is twisted with respect to the orbital plane, but the difference $\Delta\omega = \omega_{\rm d} - \omega_{\rm p}$ is rather small, $\Delta\omega \sim 5^{\rm o}$. If one assumes that the disc is thin, its inclination to the orbital plane has to be grater than $i_{\rm d} > 40^{\rm o}$. For our simulations, we choose the intermediate opening angle $\theta_{\rm d}=20^{\rm o}$ but different disc inclinations  $i_{\rm d}$ and $\omega_{\rm d}$ are investigated. Two sets of those parameters, which are generally consistent with the observations, have been investigated (see Fig.~\ref{fig5} for details of the orbits):
\begin{itemize}
  \item model I: $\omega_d=142^o,\, \Delta\omega=5^o,\, \theta_d=20^o,\,
i_d=40^o$, 
  \item model II: $\omega_d=158^o, \, \Delta\omega=20^o, \, \theta_d=20^o, \,
i_d=30^o$.
\end{itemize}
In the first model, the longitude of the equatorial plane with respect to the periastron phase is fixed to $\Delta\omega=5^o$. To achieve the agreement with periods of pulsar passage through the equatorial wind, the inclination angle has to be equal to $i_d=40^o$ (see Table~\ref{tab2}). The second model represents a more tilted disc, $\Delta\omega=20^o$. In this case, in order to cover the observed periods of the disk crossings, the inclination angle has to be $i_d=30^o$.
\begin{quote}
\begin{table}
\centering
\begin{tabular}{l l l l }
\hline 
model & \multicolumn{3}{c}{before periastron} \\
\hline
 & $t_{in}$ &$t_{out}$ & $(\varphi_1,\, \varphi_2)$  \\
\hline
observations & $\tau - 18$ &  $\tau -8$ & $(-92^o,\, -57^o)$ \\
\hline
I & $\tau - 22$ &  $\tau - 12$ & $(-100^o,\, -75^o)$ \\
II & $\tau - 15$ &  $\tau - 8$ & $(-90^o,\, -50^o)$  \\
\hline 
\hline
model & \multicolumn{3}{c}{after periastron} \\
\hline
observations & $\tau + 12$ & $\tau + 22$ & $(75^o,\, 100^o)$\\
\hline
I & $\tau + 13$ & $\tau + 25$ & $80^o,\, 105^o)$ \\
II & $\tau + 19$ & $\tau + 46$ & $(90^o,\, 130^o)$ \\
\hline
\end{tabular}
\caption{\label{tab2} The relation between the orbital phase and the time (in days) from the periastron (denoted by $\tau$). $t_{\rm in}$ and $t_{\rm out}$ are the time periods corresponding to the pulsar passage through the equatorial wind region (Connors et al. 2002) for considered models I and II.}
\end{table}
\end{quote}
\subsection{Gamma-rays from the pulsar wind region}
\begin{figure}
  \vspace{6.5truecm}
  \includegraphics{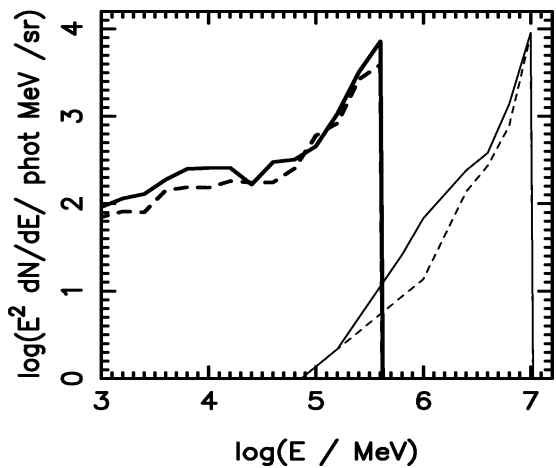}
  \caption{The differential $\gamma$-ray spectra (SED) produced in the pulsar wind zone (PWZ) for the pulsar at the periastron passage ($r = 23 R_\star$). The angle to the observer is $\theta_{obs} = 113^o$. The spectra are calculated for leptons with the mono-energetic spectra (energy $E_e = 10^7\,\rm MeV$, thin lines) and with the power law spectrum (see the text, thick lines). The distance from the pulsar to the termination shock is $r_{1} = 13\, R_{\star}$ (solid line), $r_{2} = 8.8\, R_{\star}$ (dashed line) for the geometrical models I and II.} 
\label{fig6}
\end{figure}

As we already noted, leptons, injected through the pulsar light cylinder, move radially in the pulsar wind zone. They remain almost at rest with respect to the pulsar wind. Therefore, their synchrotron energy losses can be neglected at first approximation. However, these leptons move in the radiation field of the massive star. Thus, they can produce $\gamma$-rays in the ICS process. We calculate these $\gamma$-ray spectra for two mentioned above models for the spectra of leptons and geometries of the shock (see Sect.~3.1). Note that $\gamma$-rays produced by leptons in the PWZ can be absorbed in this same soft radiation field during their propagation in the pulsar wind zone and the massive star wind region. These cascading effects are taken into account in our model as described in detail in Sierpowska \& Bednarek~(2005a). The specific $\gamma$-ray spectra escaping from the binary system were calculated for the pulsar at the periastron passage, $r = 23 R_\star$. At this location, the observer views the binary system at the angle $\theta_{obs}^{per} \approx 113^o$. The distance from the pulsar to the shock front was calculated in each model: $r_{1} = 13\, R_{\star}$ (model I), $r_{2} = 8.8\, R_{\star}$ (model II). As expected from the analysis of the optical depth, the $\gamma$-ray fluxes are larger for larger volume of the PWZ (the shock is farther from the pulsar). For monoenergetic primary leptons (model A), the spectrum of $\gamma$-rays is very flat (spectral index $\sim 0.5$).  In case of injection of primary leptons with the power law spectra (model B), $\gamma$-ray spectrum shows maximum corresponding to IC scattering of stellar photons in the KN regime. The general spectrum is significantly steeper in this case (spectral index $\sim 1.5$ below $\sim 3\times 10^4$ MeV). Similar spectra have been obtained in the case of monoenergetic leptons inside the pulsar wind nebulae (see Ball \& Kirk~2000, Bogovalov \& Aharonian~2000, Khangulyan et al.~2007). Observation of such very flat spectra from the binary system would indicate the efficient acceleration of leptons already inside the pulsar magnetosphere or in the vicinity of the pulsar light cylinder. In the case of PSR B1259/SS2883 such flat gamma-ray spectra have not been detected up to now.Therefore, the PWZ seems not to be responsible for the bulk of $\gamma$-rays escaping from this binary system. However, in other binaries, in which the pulsar wind region completely dominates the volume of the binary system, e.g. as in the case of the famous millisecond binary system PSR 1957+20, such very flat $\gamma$-ray spectra might be produced. Their observation should give important information on the spectra of primary leptons in the pulsar wind zone.

\subsection{Gamma-rays from the shock region}
\begin{figure}
  \vspace{5truecm}
  \includegraphics{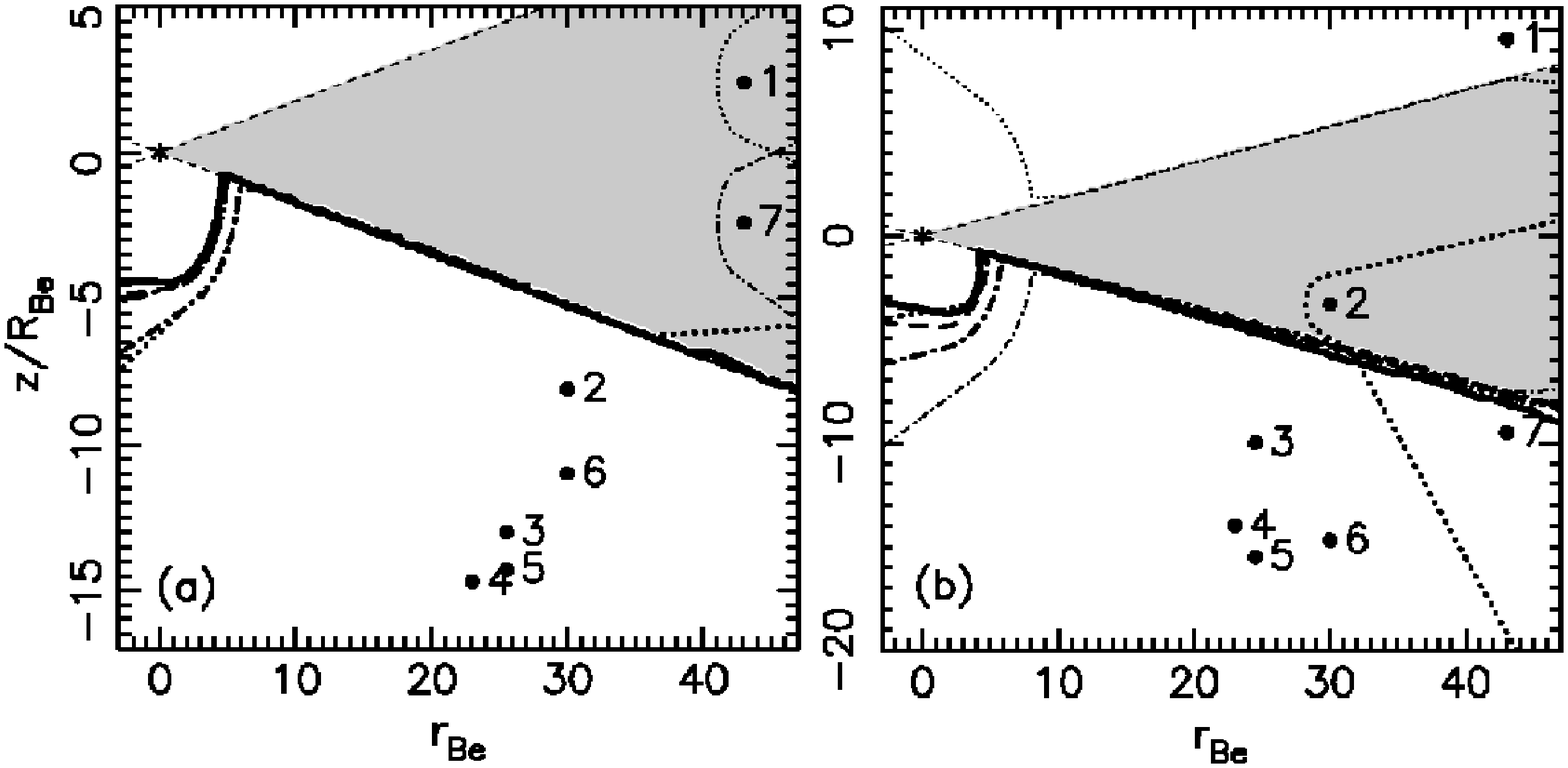}
  \caption{The geometry of  the termination shock with respect to the massive star
equatorial wind (the massive star in the center) for the subsequent pulsar
phases in the model I and II. The pulsar phases are denoted with numbers while the shock shapes are plotted with different line styles: (1) $\varphi = -90^o$ -- thin dotted line, (2) $\varphi = -60^o$ -- dotted one, (3) $\varphi = -30^o$ -- dot-dot-dot-dashed, (4) $\varphi = 0^o$ -- solid, (5) $\varphi = 30^o$ -- dashed, (6) $\varphi = 60^o$ -- dot-dashed, (7) $\varphi = 90^o$ -- thin dot-dashed.} 
\label{fig7}
\end{figure}
Since the radiation processes of leptons, accelerated at the shock region, depend significantly on the details of the shock geometry, we calculate the $\gamma$-ray spectra for two models which fulfill the general constraints concerning the positions of the pulsar close to the periastron passage. We fixed the basic parameters characterizing the wind of the massive star: the equatorial wind mass loss rate is $\dot{M}_{\rm d}=2.0\times 10^{-7} M_{\odot}$ year$^{-1}$, the polar wind $\dot{M}_{\rm p} = 10^{-3} \dot{M}_{\rm d}$, i.e consistent with the observations of this system. Other parameters of the system derived from different observations are shown in Table ~\ref{tab1}. As the orbit is highly eccentric and the plane of the equatorial wind is inclined to the orbital plane, the shape of the shock structure changes from phase to phase in a complicated way. 
\begin{figure}
  \vspace{9.truecm}
  \includegraphics{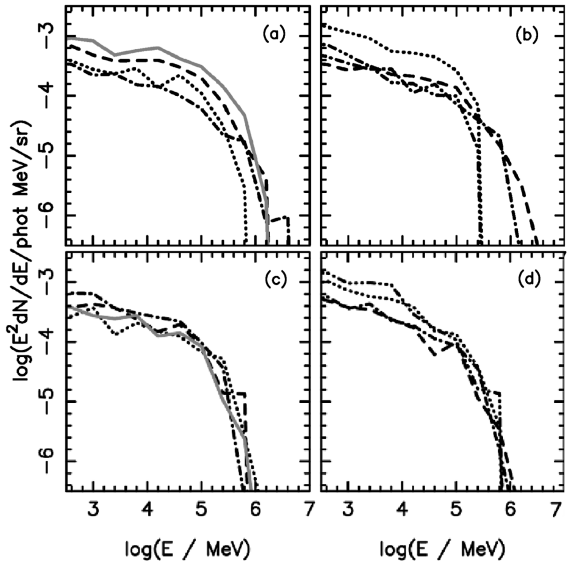}
  \caption{The differential $\gamma$-ray spectra (SED) for subsequent pulsar phases in the case of model I parameters (figures (a) and (b)) and model II ((c) and (d)). Primary leptons are injected with  the power-law spectra  and spectral index $\alpha_i = 2.5$. The $\gamma$-ray spectra are shown for different pulsar phases before periastron in figures (a) and (c): $\varphi = -90^o$ (dotted line), $\varphi = -60^o$ (dot-dashed), $\varphi = -30^o$ (dashed), $\varphi = 0^o$ (solid gray line), and after periastron in figures (b) and (d):  $\varphi = 30^o$ (dashed line), $\varphi = 60^o$ (dot-dashed), $\varphi = 90^o$ (dotted) and $\varphi = 105^o$ (dot-dot-dot-dashed). The magnetization parameter of the pulsar wind is fixed to $\sigma = 0.01$.}
\label{fig8}
\end{figure}
\begin{figure}
  \vspace{11.5truecm}
  \includegraphics{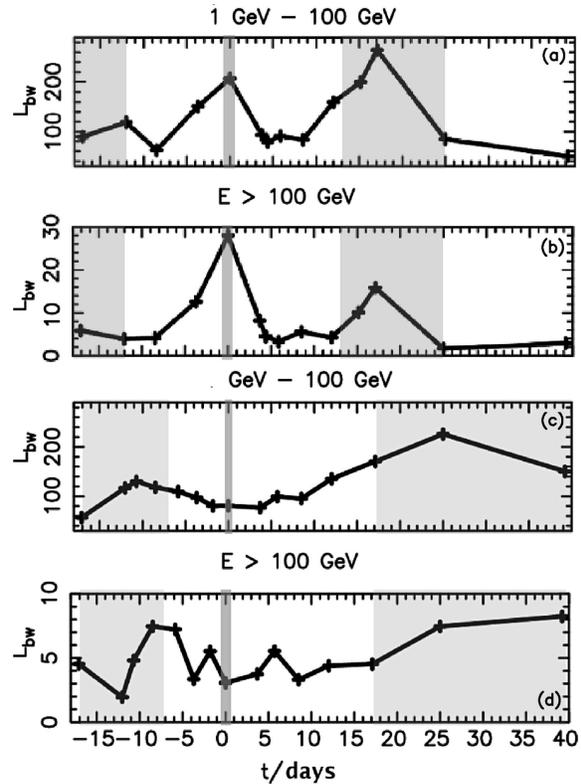}
  \caption{The gamma light curves in two energy ranges: 1 GeV - 100 GeV (a, c) and above 100 GeV (b, d). The results for the model I are presented in figures (a) and (b) while for the model II in (c) and (d). The leptons were injected with the power law spectrum and spectral index $\alpha_{\rm i}= 2.5$  (the magnetization parameter is $\sigma = 0.01$). The $\gamma$-ray luminosity, $L_{\rm bw}$, is in units of photons MeV sr$^{-1}$. The phases corresponding to the pulsar in the massive star equatorial wind are shaded. } 
\label{fig9}
\end{figure}
We apply two geometrical models mentioned above. The details of these models
are shown in Fig.~\ref{fig5} and the shapes of the shock front in subsequent phases of the binary system are shown in Fig.~\ref{fig7}. Note that for three shown phases, $\varphi = -90^o, \,  -60^o, \,  90^o$, the pulsar is immersed inside the equatorial wind.  For both models, we calculate the $\gamma$-ray spectra escaping to the observer by applying the modified Monte Carlo code developed by Sierpowska \& Bednarek (2005a). The spectra of primary leptons injected at the shock are normalized to one particle per sr.  Calculated $\gamma $-ray spectra allows us to obtain the $\gamma $-ray light curves in two energy ranges which correspond to the energy ranges of the GLAST detector ($1\, \rm GeV - 100 \, \rm GeV$) and the present Cherenkov telescopes ($>100$ GeV). We also investigate the $\gamma$-ray spectra and light curves for different power law spectra of primary leptons applying the range of spectral indexes $\alpha_i = 2.0 - 3.5$. Calculated phase dependent $\gamma $-ray spectra are shown for the spectral index of primary leptons $\alpha_i = 2.5$ (Fig.~\ref{fig8}). As expected, the spectral index of produced $\gamma$-rays is equal to $\alpha_{\gamma} \sim (2.2, \, 2.3)$ for energies $< 100 \rm GeV$. It is clearly related to the spectral index of primary leptons when the scattering occurs in the Thomson regime. $\gamma$-ray photons with larger energies are created in the Klein-Nishina regime, and so their spectra steepens. Note, that for large range of phases around periastron, the spectrum of $\gamma$-rays, calculated in terms of the model I, does not change drastically. However, if the pulsar is farther from the periastron, the cut-off in the $\gamma$-ray spectrum appears at lower energies. Note that calculations of the $\gamma$-rays spectra in such a complicated scenarios (which additionally take into account the cascade processes) are very time consuming. Therefore, the spectra suffer to some extent from low statistics. However, their main features can be clearly identified.\\
 
In Fig.~\ref{fig9} we also show the $\gamma$-ray light curves in the GLAST and Cherenkov telescopes energy ranges. The $\gamma$-ray light curves calculated in terms of the model I are characterized by clear maximum at the periastron passage and the second maximum at the phase $\tau + (15-17)$ days, ($\varphi \sim 90^o$), corresponding to the  second passage of the pulsar through the equatorial wind of the massive star.  The $\gamma$-ray light curves obtained in terms of the model II are different. They do not show the maximum at the periastron passage but only slight increase of emission at the second passage of the pulsar through the equatorial wind.\\

Note that the equatorial wind of the massive star is oriented with respect to the observer in a different way in both discussed models. The $\gamma$-rays, produced in the shock region, propagate along different paths to the observer (i.e. closer or further to the massive star), but absorption of photons can occur effectively only in the first case. The differences can be observed when we compare the photon spectra before and after periastron. Moreover, $\gamma$-ray fluxes are higher after periastron phase which is due to the fact that pulsar changes its position with respect to the observer after periastron (begins to approach the observer and the massive star is 'behind' it). Also the orientation of the shock plane changes with respect to the observer. Leptons, which propagate along the shock, are moving in direction closer to the observer line of sight. So then, produced $\gamma$-ray spectra can be observed without significant absorption having higher energies (the optical depths for pair production are the lowest when photons propagate outward the surface of the massive star). 

\section{Comparison with observations of PSR B1259-63/SS2883}
\begin{figure}
  \vspace{11.5truecm}
  \includegraphics{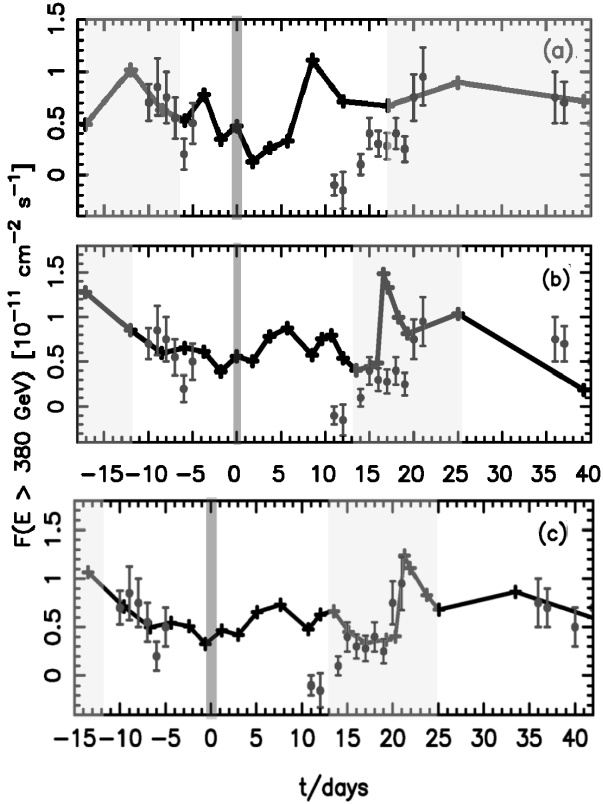}
\caption{The comparison of the $\gamma$-ray light curves obtained in the model II (a) and model I (b) with the observations of PSR B1259-63/SS2883 at energies $>380$ GeV (Aharonian et al. 2005). The primary spectrum of leptons has the spectral index $2.5$ and the magnetization parameter of the pulsar wind is $\sigma = 10^{-4}$. The periods when the pulsar crosses the massive star equatorial wind are shaded. The $\gamma$-ray light curve obtained in model I but for different orientation of the orbital plane and the plane of the equatorial wind (with $\omega_p = 148^o$) is shown in (c).} 
\label{fig10}
\end{figure}
\begin{figure}
  \vspace{9.5truecm}
  \includegraphics{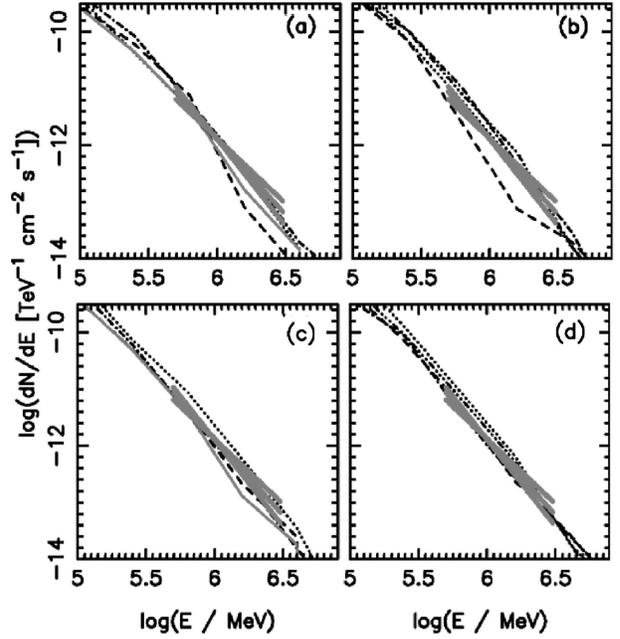}
  \caption{The differential $\gamma$-ray spectra (SED) for subsequent pulsar phases calculated in model I (with $\sigma = 10^{-4}$). The $\gamma$-ray spectrum from the binary system PSR B1259-63/SS2883 observed by the HESS telescopes  is plotted for comparison (with the errors marked by the gray lines). The spectral index of primary leptons is $\alpha_i = 2.5$. The spectra are calculated for the phases before periastron (a): $\varphi = -90^o$ (dotted line), $\varphi = -60^o$ (dot-dashed), $\varphi = -30^o$ (dashed), $\varphi = 0^o$ (solid red line),  and after periastron (b): $\varphi = 30^o$ (dashed line), $\varphi = 60^o$ (dot-dashed), $\varphi = 90^o$ (dotted), $\varphi = 105$ (dot-dot-calculated)). $\gamma$-ray spectra calculated in terms of the model II are shown in (c) and (d).}
\label{fig11}
\end{figure}

In the above discussed models I and II, we applied the value for the pulsar wind magnetization  parameter equal to $\sigma = 10^{-2}$ (as it was done in previous modelling, see Introduction). However, in such a case the energies of produced $\gamma$-rays are too low in order to be consistent with those observed from the binary system PSR B1259/SS2883 (Aharonian et al. 2005). The $\gamma$-rays with larger  energies can appear for the case of the lower magnetization parameter of the pulsar wind. Then, due to the weaker magnetic field at the shock region, leptons can be accelerated to higher maximum energies. Therefore, when comparing the result of the model with the observations of PSR 1259-63/SS2883, we apply the value $\sigma = 10^{-4}$. Moreover, we assume that  the energy spectra of primary leptons have low energy cut-offs at $E_{\rm min} = 10^5 \,\rm MeV$. We suggest that this low energy cut-off corresponds to the energies of leptons which arrive to the shock region from the pulsar. The gamma spectra and light curves are normalized to the fluxes observed from the binary system PSR B1259-63/SS2883 by the HESS experiment (Figs.~\ref{fig10} and~\ref{fig11}). The $\gamma$-ray light curve in the energy range $>$380 GeV has two distinct maxima at phases corresponding to the pulsar crossing the equatorial wind. Maximum $\gamma$-ray emission appears at phases symmetrical with respect to the periastron, $\tau \pm 17$ days ($\varphi = \pm 90^o$). The $\gamma$-ray light curve for model II (Fig.~\ref{fig10}a) has local maximum at the periastron passage and the second, stronger one, after periastron at phase $\sim \tau + 8$ days. Before the periastron the data are well fitted in case of both models.  For the investigated  model I (Fig.~\ref{fig10}b) we get a good general agreement with the HESS results, e.g. the power decreasing in period, $\tau - 11,\, \tau - 5$ days, and increasing in, $\tau + 11,\, \tau + 20$ days. A still better fit is obtained when we change by $\sim 10^{\rm o}$ the angle $\omega_{\rm p}$ which defines the position of the periastron with respect to the observer (Fig.~\ref{fig10}c). Because the observation did not cover the period $\sim \tau -5, \, \tau+10$ we cannot definitely exclude that model. 

The photon spectra for both models and $\sigma = 10^{-4}$ are compared to the HESS spectrum in Fig.~\ref{fig11}. Photon fluxes at phases after periastron ($\varphi >0$) are larger than before it. The photon spectra were calculated for different initial power law spectra of lepton. The best description of the $\gamma$-ray spectrum measured by HESS is achieved in the case of leptons injected with the power law spectrum and spectral index $\alpha_i = 2.0-2.5$. Note that for these parameters $\gamma$-ray spectra can extend up to $\sim 10 \,\rm  TeV$.  For index $\alpha_i= 2.0$ and energy range $5.5 <$log(E$_{\gamma}$)$< 6.5$ MeV,  we get the photon index is $\alpha_{\gamma} \sim  2.5$ while for $\alpha_i = 3.5$ the photon spectrum is steeper ($\alpha_{\gamma}\sim  4.0$). 

\section{Discussion and Conclusion}

We considered the model for production of high energy gamma rays in the binary system consisting of a Be type massive star and the energetic pulsar applying the details of complicated geometry of the massive star wind. The model predictions are compared with recent observations of the binary system PSR B1259-63/SS2883. In this system, due to the interaction of the pulsar wind and the massive star wind, a shock wave is created. The geometry of the shock is complex and changes with the orbital phase as the pulsar passes through equatorial or polar wind of the massive star. Such structure of the massive star wind is characteristic for a Be type star. In the considered system, the equatorial wind plane is inclined to the orbital one. $\gamma$-rays are produced by leptons which are accelerated at the shock region and after that scatter thermal photons from the massive companion. Since the soft photon field is anisotropic, we take into account the details of the geometry of the binary system. The change of the separation between the stars with the phase of the binary and the complex structure of the massive star wind have also big impact on the conditions for acceleration of leptons and their interactions with the stellar radiation. The maximum energies of leptons and their energy losses are determined by the magnetic field strength at the shock region which in turn is defined by the magnetization parameter ($\sigma$) of the pulsar wind. In our calculations a value of $\sigma = 10^{-2}$ and $10^{-4}$ is considered. The presented model takes into account the complicated geometry of the system and still it has to depend on many parameters which describe the position of the orbital plane and the equatorial plane of the stellar wind with respect to the observer.

In such a complicated geometrically  model, leptons can be accelerated in some cases relatively close to the stellar surface. Therefore, the possible absorption of the $\gamma$-rays produced in the IC scattering process has to be also taken into account.  The optical depths for leptons on ICS within the pulsar wind region are clearly too low for effective production of $\gamma$-ray photons. However, leptons, which arrive from the pulsar to the shock region, are trapped by turbulent magnetic field after the shock and can be additionally accelerated. If the shock region is relatively close to the massive star, IC scattering becomes important and high energy $\gamma$-rays are produced effectively.

We calculate the $\gamma$-ray spectra produced in the shock region for power law spectra of leptons in two different geometrical models. By applying the sets of parameters, I: $\Delta\omega_d=5^o, \, \theta_d=20^o, \, i_d=40^o$ and II: $\Delta\omega_d=20^o, \, \theta_d=20^o, \, i_d=30^o$, we reach the general consistency with the TeV $\gamma$-ray observations. The best fit of the gamma light curve measured by the HESS telescopes (Aharonian, et al.~2005) is obtained by applying the model I with $\sigma=10^{-4}$. However, this fit has been reached assuming that the periastron position of the pulsar is defined by the angle $\omega_{\rm p} = 148^o$ contrary to the present estimates, i.e.  $\omega_{\rm p}\sim 138^o$. Note, however that the estimation of the angle $\omega_{\rm p}$ depends on massive star wind parameters such as the inclination of the disc (applied from Wex et al. 1998) and the mass loss rate, which in case of PSR B1259-63 was fixed to $5\times 10^{-8}$ M$_\odot$yr$^{-1}$, and which are in fact not very well known.

The calculated $\gamma$-ray spectra are in agreement with the TeV HESS data provided that the spectral index of the spectrum of leptons is $\alpha_i = -2.5$. The slope and also the fluxes of the photon spectra do not change with orbital phase, as was reported by the HESS collaboration.  The final light curve for model I ($\sigma=10^{-4}, \omega = 148^o$) is characterized by a maximum for the phases when the pulsar crosses the equatorial disc after periastron. This fit differs from the observational data for phases when pulsar approaches the equatorial disc. This can be due to the complex geometry and structure of the shock region when the pulsar wind interacts with the strong equatorial wind. The model assumes that the acceleration of leptons occurs locally at the shock. In reality, this acceleration process of leptons arriving to the specific place at the shock can extend to other parts of the shock. Such a non-local acceleration process is very difficult to consider, but it might be important and influence some details of the calculated $\gamma$-ray light curves. 

Based on the reported calculations, we conclude that the general shapes of the TeV and GeV $\gamma$-ray light curves close to the periastron should be quite similar. However, farther out from the periastron, the GeV emission should continue with larger flux. This feature can be tested by the future observations in the GeV energy range by the AGILE and GLAST telescopes.

\section*{Acknowledgments}
This work is supported by the Polish MNiI grant No. 1P03D01028.

\end{document}